# Quantitative imaging of carbon dimer precursor for nanomaterial synthesis in the carbon arc

V. Vekselman, A. Khrabry, I. Kaganovich, B. Stratton and Y. Raitses

Princeton Plasma Physics Laboratory

Princeton, NJ 08543, USA

Delineating the dominant processes responsible for nanomaterial synthesis in a plasma environment requires measurements of the precursor species contributing to the growth of nanostructures. We performed comprehensive measurements of spatial and temporal profiles of carbon dimers ($C_2$) in atmospheric-pressure carbon arc by laser-induced fluorescence. Measured spatial profiles of $C_2$ coincide with the growth region of carbon nanotubes [1] and vary depending on the arc operation mode, which is determined by the discharge current and the ablation rate of the graphite anode. The $C_2$ density profile exhibits large spatial and time variations due to motion of the arc core. A comparison of the experimental data with the simulation results of self-consistent arc modeling shows a good agreement. The model predicts well the main processes determining spatial profiles of carbon dimers ($C_2$).

## 1. Introduction

Understanding the synthesis of nanomaterials by an atmospheric pressure arc discharge such as a carbon arc for synthesis of fullerenes, nanotubes, and graphene [2-4] requires knowledge of the plasma parameters and composition in the region of nanostructure formation and growth. There is currently no satisfactory theoretical description of this process due to the lack of reliable and accurate measurements. It is challenging to perform *in-situ*, non-invasive measurements in the harsh arc discharge environment. *Ex-situ* evaluation of synthesized nanostructures does not provide detailed understanding of the growth mechanism and conditions. The use of probes for detection of densities, temperatures, and other environmental parameters or collection of samples has the drawback of inducing perturbations in the region of interest.

Starting from the discovery of synthesis of carbon nanotubes (CNT) in a carbon arc by Iijima [5] a variety of methods for producing carbonaceous nanostructures have been developed [6-8]. In parallel, *in-situ* measurements of synthesis precursors were carried out using optical techniques. Lange, *et al* [9] applied optical emission spectroscopy (OES) to study arc plasma composition and diatomic carbon ($C_2$) radicals; this work was continued in [10-12]. Other OES measurements of carbon species in arcs were carried out in [13,14]. However, OES measurements are line-integrated along the observation sightline and additional assumptions are required to deduce density and temperature profiles. These assumptions introduce ambiguity in the interpretation of OES measurements. Advanced laser-based spectroscopy was applied to study carbon species in the laser vaporization synthesis method [15-17].

Carbon atoms in arc undergo several transformations before incorporating into a nanostructure. They are evaporated from the anode material in form of molecules ($C_2$, $C_3$) [18] and larger particles [19] and then dissociated in the hot arc core to carbon atoms. At lower temperatures (3000-5000

K) carbon atoms associate back to $C_2$ and $C_3$ molecules [20], as evidenced by intense emission of molecular bands from these species. Recent studies show that the $C_2$ content may affect the properties of the synthesized nanomaterial [21,22]. Abundance and the opportunity to affect the synthesis path reveal the critical role of $C_2$ as an important synthesis precursor.

Laser Induced Fluorescence (LIF) differs from other spectroscopic techniques due to its high selectivity, sensitivity and high spatial and temporal resolution. LIF is inherently free from the effects of laser background light, which is a significant advantage for application to high pressure plasmas in which scattered laser light is very intense. Atmospheric pressure arc plasmas containing molecular species are characterized by complex chemistry that depends on the plasma state. The volume occupied by plasma is relatively small and features large time-evolving gradients. Thus, the LIF technique is best suited to the study of carbon precursor species participating in synthesis. A variant of LIF, known as planar LIF, allows instantaneous measurement of density distribution profiles with all the benefits of conventional LIF. In planar LIF, the laser beam is shaped into a sheet geometry and a two-dimensional detector is used. LIF and planar LIF techniques are described in detail in [23-25].

Here we report the first comprehensive measurements of spatial and temporal profiles of carbon dimers in atmospheric-pressure carbon arc by the laser-induced fluorescence technique. Planar LIF was applied to obtain quantitative density distributions of $C_2$ in different modes of arc operation. The spatial profiles of $C_2$ forming a bubble-like shape [26] were found to coincide with growth region of CNTs [1]. The $C_2$ density profiles reported in this paper are therefore of great interest for modeling of synthesis processes in the carbon arc.

### 1.1. Description of quantitative planar LIF

The rate equation for the population of the excited state $n_2$ can be written as

$$\frac{dn_2}{dt} = \frac{B_{12}}{c} n_1 I_\nu - \frac{n_2}{\tau_{eff}} \quad (1)$$

where $n_1$ is the population of the lower (pumping) state, $B_{12}$ is the Einstein absorption coefficient and $\tau_{eff}$ is the effective lifetime of the excited state which accounts for all processes leading to depopulation of the excited state excluding fluorescence. The laser stimulated depopulation term is neglected here assuming $n_1 \gg n_2$. $I_\nu$ is the spectral laser intensity defined as

$$I_\nu = \frac{E_L}{A_L \tau_L \Delta\nu_L} \Gamma \quad (2)$$

where $E_L$ is the laser energy per pulse and $\Gamma$ is the dimensionless lineshape integral accounting for the interaction between the laser and transition absorption lineshape profiles, $\Gamma = \int a(\nu)g(\nu)d\nu$. Here $a(\nu)$ is the absorption lineshape centered at transition frequency and normalized by unity, $\int a(\nu)d\nu = 1$; $g(\nu)$ is the laser lineshape normalized to unity, $\int g(\nu)d\nu = 1$. A typical approximation of $\Gamma \sim 1$ reflects inequality $\Delta\nu_L \gg \Delta\nu_{tr}$ where the $\Delta\nu_{tr}$ is the transition linewidth.

The measured LIF signal $S_{LIF}$ is proportional to the number of fluorescence photons

$$S_{LIF} = n_2 A_{21} \tau_{eff} V \frac{\Omega}{4\pi} K \quad (3)$$

where $A_{21}$ is the Einstein coefficient for spontaneous emission from the excited state, V is the interaction volume, Ω is the solid angle observed by the collection optics and K is a cumulative coefficient related to the properties of the detector and collection optics and the spatial profile of the laser intensity. Since the $n_1$ can be correlated with the population of the ground state through the Boltzmann factor, $n_1 = n_0 f_B$, the ground state density can be obtained from the measured LIF signal without accounting for saturation effect as

$$n_0 = S_{LIF} \cdot \left[ f_B \frac{B_{12}}{c} \frac{E_L}{A_L \Delta \nu_L} \Gamma \cdot A_{21} \tau_{eff} V \frac{\Omega}{4\pi} K \right]^{-1} \quad (4)$$

Direct implementation of Eq. 4 requires knowledge of cumulative parameter K, interaction volume and observation solid angle which are hard to obtain. Therefore, these undetermined parameters are commonly estimated by calibration via Rayleigh or Raman scattering. In Appendix B, we describe in detail how these parameters were obtained by Rayleigh scattering measurements.

Phenomenon of transition saturation is important to consider in experiments with single shot LIF measurements. Balancing between signal-to-noise ratio (SNR) and linear LIF signal response shifts the level of the laser intensity towards the saturation region. In multi-shot LIF measurements the accumulation of the signal allows lowering the laser intensity keeping SNR the same. Planar LIF geometry introduces spatial dependence of the laser intensity in transverse direction (relative to the laser beam propagation); it is described in Appendix B. Therefore, the saturation effect becomes, in general, space dependent; it is discussed in Appendix D.

Another important phenomenon related to the LIF measurements in atmospheric plasmas is the shortening of the excited state lifetime. It is mostly due to quenching by surrounding atoms and molecules, plasma electrons and branching channels. In synthesis arc, participation of the species of interest in synthesis processes contributes to the lifetime shortening as well. Therefore, time-resolved LIF measurement of the excited state lifetime is required (Appendix C). It also allows fine tuning of the fluorescence acquisition time in planar LIF to further improve SNR.

## 2. Experimental setup

### 2.1. Arc setup

The arc system used in this experiment is an atmospheric pressure DC arc discharge that is described in detail in [26]. It consists of a stainless steel 6-way cross equipped with four optical viewports allowing access to the inter-electrode region where the plasma is formed. Two graphite electrodes of 9.5 mm (cathode) and 6.5 mm (anode) in diameter are aligned vertically and their position can be adjusted via two computer controlled motorized translational stages. Helium was used as the buffer gas. The flow of helium was regulated by a feedback controlled vacuum line and flow needle valve via a PID algorithm (using LabVIEW) to maintain 66.6 kPa total pressure in the chamber. The arc setup and LIF diagnostic assembly are schematically shown in Figure 1.

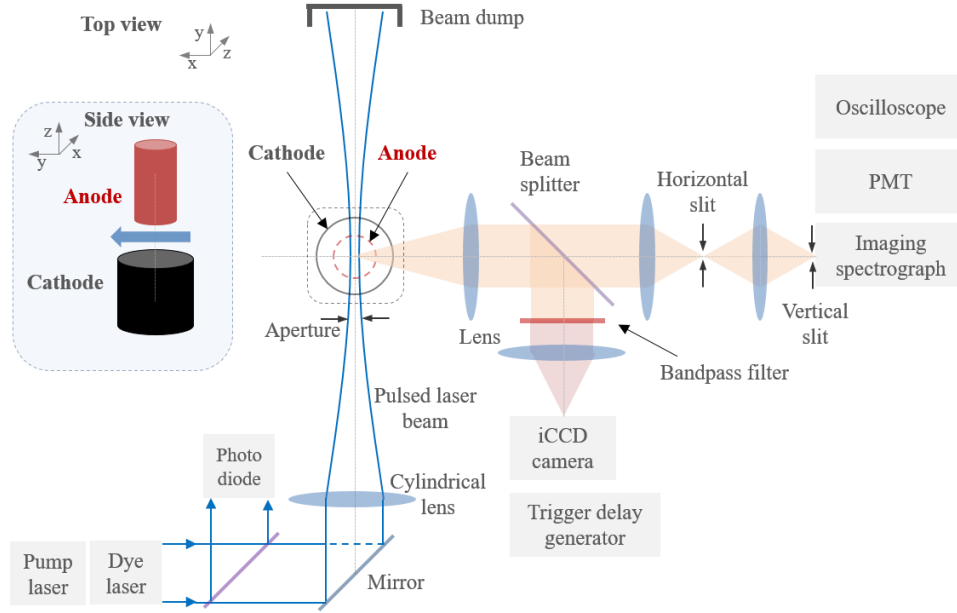

Figure 1. Schematic of the experimental setup and LIF diagnostics.

The arc between graphite electrodes was initiated by bringing the electrodes into contact and then moving them apart. The power supply maintained a constant discharge current while the discharge voltage was determined by the inter-electrode gap length. There are two modes of arc operation that differ by the anode ablation rate [26]. In the low ablation mode, the anode material ablation rate is below 1 mg/s and almost independent of the arc current. In contrast, the high ablation mode features larger anode ablation rates (up to 10 mg/s) and strong correlation with the current. Recent study has also revealed dependence of the ablation rate on the gap length at fixed current. The transition between low and high ablation mode occurred at about 55 A current for a given electrode geometry.

The two arc operation modes are also characterized by different arc behavior. Arc core motion is confined between the electrodes in the low ablation mode and oscillates with frequencies up to 10 kHz. In high ablation mode, the arc core attachment extends to the side surface of the anode and features additional oscillations that are dominant in the frequency range below kHz [27]. Such unpredictable dynamics of the arc complicates the application of the diagnostics. In the current paper we present results for discharge currents of 50 A and 60 A corresponding to the low and high ablation modes of the carbon arc.

### 2.2. Planar LIF setup

The planar LIF diagnostic setup was employed to obtain instantaneous density distribution profiles. The LIF laser system consisted of a Spectra-Physics (Sirah) Cobra-Stretch dye laser pumped by the third harmonic (355 nm) of a Continuum SL-III Nd:YAG laser. For excitation of the $C_2$ (3,1) transition used in this study, Coumarin 440 laser dye dissolved in ethanol was used. The dye laser was operated near 437 nm with a linewidth of 0.06 cm$^{-1}$. The fluorescence was detected near 470 nm. The output energy of the pump laser was approximately 100 mJ and the dye laser output energy was about 10 mJ, with an estimated energy fluctuation of 12%. The dye laser beam passed through a variable attenuator (1/2 wave-plate and linear polarizer mounted on a rotational stage) which allowed the laser energy to be varied continuously. The laser beam was then directed into an adjustable aperture installed in front of the chamber entrance quartz window.

Planar LIF utilizes a sheet-like laser beam profile oriented vertically along the axis of the arc electrodes. The beam shaping was achieved by a cylindrical lens as shown in Figure 1. The thickness of the laser beam was estimated to be 50 μm using the Rayleigh criterion. The height of the laser beam was limited by the aperture. Due to the laser shaping, the transverse profile of the laser beam intensity shows a gradual decrease toward the edges (see Appendix). This implies a need for a laser power distribution correction for the LIF measurements in planar geometry.

An intensified CCD (iCCD, Andor iStar) camera was used to obtain the LIF signal from the sheet-like interaction volume in one laser shot. The iCCD camera was coupled to an objective lens and bandpass filter (Andover 470FS10-50, centered at 470 nm with bandwidth 10 nm). The spatial resolution was 1/36 mm per pixel. The exposure time of the iCCD camera was adjusted to match the duration of the fluorescence signal (see Appendix) to minimize the collection of radiation from the plasma volume passing through the bandpass filter. The background was subtracted from the collected signal. However, this subtraction is a key contributor to the measurement error due to the long time intervals between subsequent acquisitions with the laser on and off.

## 3. Results and discussions

### 3.1. Experimental results

A set of frames showing the measured $C_2$ density distribution during the first half minute of the arc operated at 50 A current is shown in Figure 2. This current corresponds to the low ablation mode of the arc. The arc voltage and current waveforms are shown in the bottom-left with time markers corresponding to recorded planar LIF images (1)-(4). After initiation of the arc $C_2$ always dominates in the region near the anode surface ablated by the arc core (Figure 2). However, flare-like structures of $C_2$ extending towards the cathode can be observed during the first several seconds as seen in Figure 2(1) and (3). Typical $C_2$ densities in these flares are in the range $10^{16}$ -$10^{17}$ cm$^{-3}$. At these times, the $C_2$ distribution is non-uniform and asymmetric relative to the electrode axis. It is unclear if $C_2$ is formed in the anode vicinity due to evaporation from micro-particles [19] and large molecules dissociation [18] or from the carbon atom flux from the evaporated anode.

Later in time, corresponding to images (3) and (4), the bubble-like structure of the $C_2$ distribution is formed and remains until the arc extinction. In this regime, the largest density of $C_2$ up to $3 \cdot 10^{16}$ cm$^{-3}$ is observed near the anode surface. In the bubble boundary layer, the density is in the $10^{15}$ cm$^{-3}$ range. We note that planar LIF images are generated from the intersection of the laser sheet beam with $C_2$ volumetric distribution and therefore the observable shape of the bubble is affected by the bubble position. Symmetrical $C_2$ bubble shapes, similar to those shown in Figure 2(4), were typically observed after 10-20 s of arc operation. The size of the $C_2$ bubble in the radial direction is larger for larger gaps $l_{gap}$ and typically exceeds the anode size when $l_{gap} \geq r_a$, where $r_a$ is the anode radius.

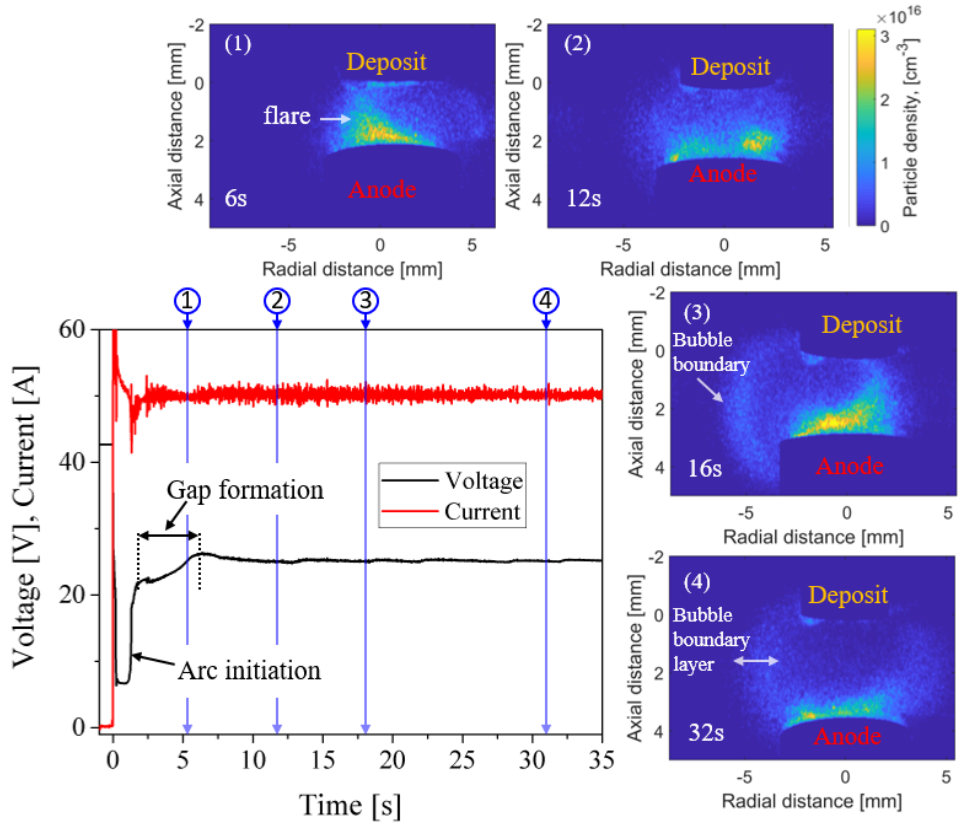

Figure 2. Evolution of the $C_2$ density distribution in arc in low ablation mode (current 50 A). Time is increasing from (1) to (4) and shown in the bottom-left corner. Final inter-electrode gap length is about 3 mm. Anode 6.5 mm in diameter, cathode – 9.5 mm.

At small gaps ($l_{gap} < r_a$), the $C_2$ bubble is confined between electrodes and exhibits sporadic motion, see Figure 3. As was shown by the fast frame imaging diagnostic in [26] the bubble follows the arc core dynamics. In Figure 3 the arc is operated in high ablation mode and $C_2$ densities in the bubble boundary layer are approaching few $10^{16}$ cm$^{-3}$. The $C_2$ density near the anode surface is up to $10^{17}$ cm$^{-3}$. The $C_2$ concentration increase is mainly due to the increase of the current and partially due to the decrease of the gap length.

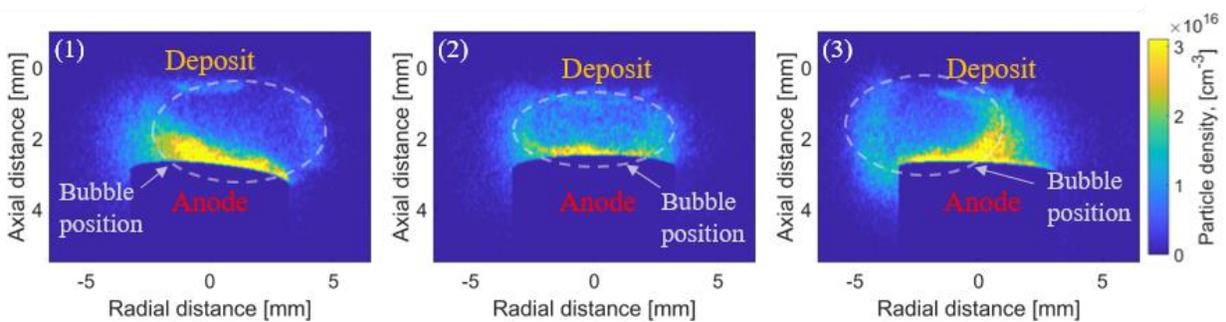

Figure 3. The $C_2$ density distribution in arc discharge in high ablation mode (current 60 A) showing the bubble motion. Frames are selected arbitrary during the arc run. Inter-electrode gap length is about 2 mm. Anode 6.5 mm in diameter, cathode – 9.5 mm.

The size of the $C_2$ bubble is smaller for shorter gaps as it is clearly seen by comparing Figure 2(4) and Figure 3(2). However, in both cases of small and large gaps, the bubble boundary extends up to a distance of 3-5 mm in the radial direction (compare Figure 2(4) and Figure 3(3) or Figure 3(1)). In high ablation mode, the presence of $C_2$ at 5 mm radial distance is due to the bubble motion (e.g., the bubble shifts off the electrode axes). This is an important observation because recently applied fast movable witness plate diagnostics detected CNTs at radial distances 3-11 mm [1]. Thus, from the synthesis point of view, $C_2$ can be present in the growth region following two different scenarios corresponding to low and high ablation modes:

1. with concentration about $10^{15}$ cm$^{-3}$ and small variation during the arc operation (low ablation mode)
2. with concentration above $10^{16}$ cm$^{-3}$ but time-dependent, determined by the bubble (arc) motion (high ablation mode).

The highest yield of synthesis of carbon nanomaterials (e.g., nanotubes) is typically obtained in arcs operated in high ablation mode with short inter-electrode gaps. This is well correlated with observation of the largest concentration of the carbon molecular precursor ($C_2$) in this regime. Therefore, the $C_2$ density values and time evolution in synthesis region are important inputs for synthesis modeling. It follows from the arc modeling described in the next section, that the $C_2$ content is high in the synthesis region (temperature 2000-4000 K) where catalyst particles condense into droplets and CNTs grow.

### 3.2. Modeling

The measurements presented in this work are complemented by 2D-axisymmetric steady state simulations of carbon arc discharge in helium atmosphere performed with the computational fluid dynamics (CFD) code ANSYS CFX which was customized for this purpose. A fluid model of the plasma was coupled with models of heat transfer and current flow in the electrodes to allow self-consistent determination of temperature and current density profile profiles, non-uniform ablation, and deposition of carbon at surfaces of the electrodes. Non-equilibrium effects in the plasma, such as thermal and ionization non-equilibrium, electron diffusion, thermal diffusion and charge separation in near-electrode sheaths were accounted for more accurate description of the plasma-electrode interaction. This enables better prediction of ablation and deposition areas and temperature profiles to determine the distribution of carbon species in the plasma volume. Transport coefficient in plasma were taken from [28]. The equilibrium chemical composition of carbon species was assumed to be identical to [20]. The plasma model was benchmarked [29] against previous numerical studies [28]. Detailed description of the 2D simulations and produced results will be published in a separate paper.

The result of a 2D simulation showing the density distribution of $C_2$ in the arc is presented in Figure 4(a). The input parameters for the model are the same as in the experiment described in the preceding sections of this paper (graphite electrodes, 6.5 mm anode and 9.5 mm cathode in diameter, 66.7 kPa helium gas pressure, current 50 A). However, while the carbon fluxes towards the cathode were accounted for in the model, the geometry change due to the growth of the cathode deposit was not simulated, i.e., the cathode surface remains flat in the simulation in contrast to deposit formation in experiment, as shown in Figure 4. The obtained C₂ density distribution shown in Figure 4(a)) features the bubble-like shape similar to the shape seen in the measurements shown in Figure 4(b). The bubble size, bubble boundary layer thickness and density obtained from the simulation are in good agreement with the experiment. In addition, the simulation showed extensive formation of C₂ near the anode surface as observed in the experiment.

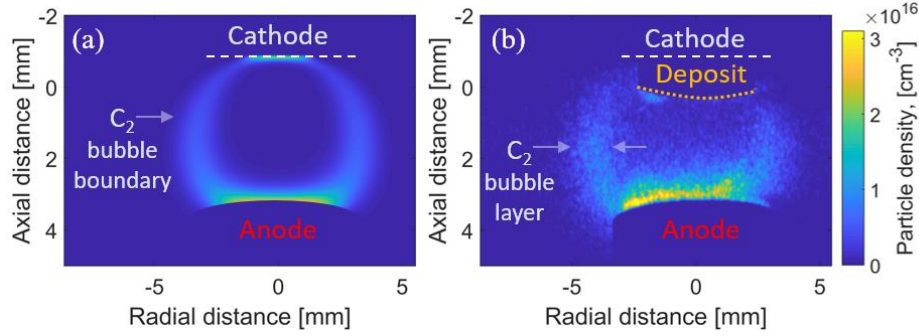

Figure 4. Comparison of (a) simulated and (b) measured C₂ density distribution in the arc at current 50 A. change in the cathode surface topology due to the growth of deposit (b) was not accounted for in the simulation (a).

The density of C₂ molecules in the bubble boundary layer can also be estimated from the transport equation for carbon atoms. According to the simulations and measurements, carbon is present only in atomic form in the arc core, where temperature is about 0.9 eV [26]. Approaching the arc periphery, the temperature decreases and reaches a range at which C₂ molecules are significant fraction of the composition of the carbon species mixture; this region corresponds to the bubble boundary layer further denoted as the bubble layer, see Figure 4 for notation. According to [20], this temperature range is about 4000 K – 5000 K; at lower temperatures, larger carbon molecules are the primary species present. According to this general picture, carbon atoms diffuse from the arc core through the bubble layer where they associate and form carbon molecules. Note that according to the simulations, in the bubble layer helium atoms are dominant species, because this region is sufficiently far from the anode ablation area. The total flux of carbon atoms through the entire bubble can be estimated as:

$$\Gamma_{C,out} = 2\pi R_{bubble} d_{arc} \cdot D_{C-He} \frac{n_{C,in}}{L_{bubble}}, \quad (5)$$

where $R_{bubble}$ is the radius of the bubble, $d_{arc}$ is the inter-electrode gap, $L_{bubble}$ – thickness of the bubble layer, $n_{C,in}$ is the carbon density at the inner side of the bubble boundary, $D_{C-He}$ is the diffusion coefficient of carbon in helium, which can be expressed as:

$$D_{C-He} \approx \frac{1}{3} \frac{v_{th,C}}{n_{He} \sigma_{C-He}}. \quad (6)$$

Here, $\sigma_{C-He} \approx 3 \cdot 10^{-19} m^2$ [30] is collisional cross-section, $n_{He}$ is helium number density, $v_{th,C}$ is thermal velocity of carbon atoms:

$$v_{th,C} = \sqrt{\frac{8kT}{\pi m_C}}, \quad (7)$$

where $k$ is the Boltzmann constant, $T$ is the gas temperature in the bubble layer, and $m_C$ is the carbon atom mass. The helium number density can be assessed from:

$$n_{He} \approx \frac{p}{kT}, \quad (8)$$

where $p$ is the total background gas pressure.

The total flux of carbon atoms escaping from the arc core can be also determined from the difference between measured rates of carbon ablation from the anode surface and its deposition on the cathode $\Gamma_{C,out}$, which is about 0.2 mg/s [26]

$$\Gamma_{C,out} = (G_{ablation} - G_{deposition})/m_C. \quad (9)$$

Knowledge of the carbon flux escaping from the arc allows us to estimate the carbon density at inner side of the bubble boundary, $n_{C,in}$, from Eq. (5) and then express number density of $C_2$ molecules. According to [20], density of $C_2$ molecules is roughly equal to density of C atoms in the given temperature range. Thus, the average density of C atoms is roughly a half of $n_{C,in}$:

$$n_{C_2,bubble} \approx n_C \approx 0.5 n_{C,in}. \quad (10)$$

Note that ratio $R_{bubble}$ to $L_{bubble}$ in Eq. (5) is defined by radial temperature profile in the arc, which is rather similar in most carbon arcs and is about 3.

Finally, from Eqs. (5) – (10) it follows that:

$$n_{C_2,bubble} \approx \frac{G_{ablation} - G_{deposition}}{12\pi m_C D_{C-He} d_{arc}}. \quad (11)$$

For the arc with inter-electrode gap length of 4 mm Eq. (11) predicts $C_2$ density is about $10^{16} cm^{-3}$ which is within the range of simulated and measured results.

## 4. Implications for understanding the synthesis processes

The results of this study have several important implications for understanding the synthesis of nanomaterials in a carbon arc:

a) The $C_2$ bubble boundary layer is located in the reported single-wall CNT growth region. According to our simulations and [20], the formation of $C_2$ is followed by the formation of larger carbon molecules, $C_3$, $C_4$, *etc.* The recent study [22] shows that presence of the $C_2$ can accelerate nucleation processes in a CVD synthesis of carbon nanotubes. Therefore, molecular carbon plays an important role in the synthesis process and should be considered in synthesis simulations.

b) The dynamics of the $C_2$ bubble, mostly populated in the boundary layer, introduce an additional time dependence in the synthesis path. Synthesis in carbon arcs, routinely operated in high ablation mode, is modulated due to the arc dynamics that has not been considered in synthesis simulations. Although the intrinsic time scale of nanostructure growth is believed much shorter than the μs time scale of the arc oscillations, it can affect the properties of the synthesized nanomaterial due to non-intentional functionalization in such a time varying plasma environment. This follows from comparison of the ms-time scale of the nanostructure movement through the bubble layer and the tens of μs characteristic arc oscillation time.

## 5. Conclusions

Quantitative measurement of the spatial profiles of dominant precursor species participating in nanomaterial synthesis is critical for understanding the processes leading to growth of nanostructures such as carbon nanotubes in the carbon arc plasma. We presented the results of a quantitative study of carbon dimer ($C_2$) spatial profiles by the planar Laser Induced Fluorescence (LIF) technique. This technique is applied for the first time to a carbon arc for nanomaterial synthesis at atmospheric pressure. The results not only confirm the recently observed [26,27] bubble-like structure of $C_2$ in the carbon arc which coincides with the growth region of carbon nanotubes [1], but also provides absolute values for the densities of molecular carbon species.

The highest $C_2$ density is observed near the surface of the graphite anode consumed by the arc evaporation, $10^{16}$-$10^{17}$ cm$^{-3}$. In the volume of the inter-electrode gap, the $C_2$ spatial profile has a distinguished bubble-like shape with densities in the range of $10^{15}$-$10^{16}$ cm$^{-3}$. This shape is formed shortly after the arc initiation and strongly affected by the length of the inter-electrode gap. The measured profiles of $C_2$ densities are in good agreement with simulation results.

It was found that different conditions form in the growth region of nanostructures for two arc ablation modes. In the high ablation mode, the $C_2$ concentration is in $10^{16}$ cm$^{-3}$ range, but the bubble-like spatial profile undergoes a sporadic motion following the arc core dynamics. In contrast, in arcs operated in the low ablation mode, the $C_2$ concentration in the bubble boundary layer is low, in the $10^{15}$ cm$^{-3}$ range, and the bubble does not move largely during the arc duration. However, in both arc modes the location of the $C_2$ bubble layer (3-5 mm from the electrode center) coincides with reported region of carbon nanotube growth [1]. The results of this study are important for further development of predictive models of nanomaterial synthesis in carbon arcs.

## 6. Acknowledgments

The authors are grateful to A. Merzhevsky (electrical engineer, PPPL) and T. Huang (visiting student from Soochow university) for support with assembly and calibration of the experimental setup, and to Dr. M. Shneider (Princeto University), Dr. A. Gerakis (PPPL), Dr. S. Yatom (PPPL), Dr. V. Nemchinsky and Dr. A. Khodak (PPPL) for fruitful discussions.

The development of the LIF setup and the absolute density calibration was supported by the U.S. Department of Energy, Office of Science, Basic Energy Sciences, Materials Sciences and Engineering

Division. Application of LIF diagnostics to the carbon arc and arc modeling was supported by the U.S. Department of Energy (DOE), Office of Science, Fusion Energy Sciences.

The digital data for this paper can be found at
http://arks.princeton.edu/ark:/88435/dsp01x920g025r.

# Appendix

This appendix, describes the steps carried out to calibrate the LIF setup and to process planar LIF images to obtaining quantitative measurements of the $C_2$ density in the arc discharge. It includes a description of LIF tuning to the (3,1) transition, Rayleigh scattering measurements to obtain calibration constants of the LIF detection system, evaluation of the saturation effect, and determination of the effective lifetime of the excited state.

### A. LIF tuning

Two procedures in the calibration of the planar LIF setup were carried out using a conventional LIF setup. First, we describe the tuning of the LIF laser to the $C_2$ (3,1) transition wavelength. In conventional LIF measurements the fluorescence radiation from the interaction volume was imaged by an f/4 optical system onto the entrance slit of a spectrometer (Horiba iHR550 with 2400 gr/mm grating). The light-collection system optical axis was perpendicular to the laser beam. The image formed by the collection lens was spatially filtered using two adjustable slits, defining the probed interaction volume as the intersection of the laser beam and observation cone as shown in Figure 1. The photo-multiplier tube (PMT, Hamamatsu R1894) was coupled to the output of the spectrometer, which was equipped with an additional motorized slit to control the spectral bandwidth of the light entering the detector. As a result, undesired light (stray light, scattered radiation from optical elements, laser beam and especially plasma emission) were effectively suppressed by a 0.4 nm spectral window.

The pump laser was controlled by a trigger generator (BNC 575). Signals from the PMT and the photodiode were recorded by a 1 GHz oscilloscope (Lecroy WaveSurfer 10). To ensure reliable detection of the LIF signal the oscilloscope was triggered by a reference TTL signal generated by the trigger generator, accounting for the time delay of the laser beam output (Q-switch and delays in internal electrical circuits) and propagation time. This approach also ensures the background signal recording during desired time interval.

The laser beam was expanded along a distance of 2 m to form a uniform beam profile at the aperture plane. This served to diminish the possibility of local saturation by laser spots ruining the LIF measurements. During the LIF measurements the laser beam energy was kept below 50 µJ. Laser energies were monitored using a photodiode that was calibrated by an energy meter (Thorlabs PM121D) with a pyroelectric head (Thorlabs ES220C).

As reported in previous work [26], the Swan bands ($a\ ^3\Pi_u - d\ ^3\Pi_g$) of $C_2$ are major contributors to the arc emission in the visible wavelength range. The Swan (3-1) band was used for measurements of $C_2$ density distribution in this work.

Figure 5 shows a LIF excitation scan around the (3,1) bandhead.

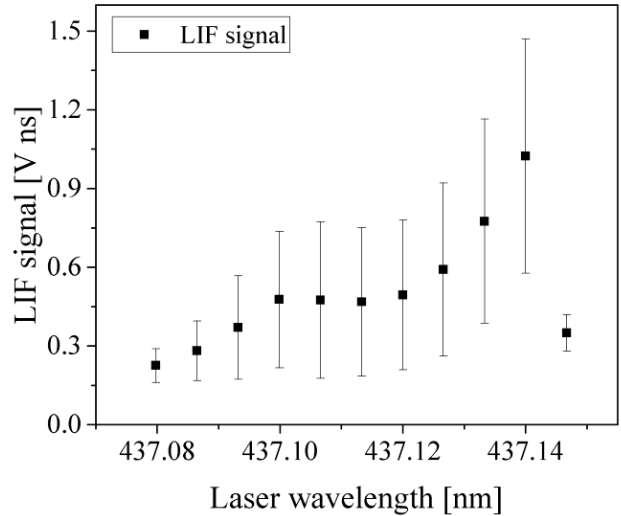

Figure 5. Laser wavelength scan near the (3,1) bandhead of the $C_2$ Swan band. The fluorescence is collected near 469 nm. The arc discharge current was 55 A.

The band structure was not observed due to the limited duration of the arc operation and the time- and space- variation of the $C_2$ distribution in the arc. The region of $C_2$ formation, which partially correlates with the emission pattern and has a bubble-like shape, evolves in space (mm scale) and time (kHz range) due to motion of the arc core [26]. Therefore, the probe volume, which is fixed in space, contains carbon dimers in different states for each subsequent laser shot, which are separated by a 0.1 s interval. This difficulty can be partially overcome by cross-correlation of the $C_2$ bubble position with the laser shots and by operating the arc in low ablation mode. In this regime, the $C_2$ bubble fluctuations are strongly suppressed and behave like so-called breathing oscillations where the bubble center of mass remains fixed in space. Together with statistical averaging of the LIF signals the obtained excitation spectrum was only used for the tuning of the laser wavelength to the (3,1) transition. The $C_2$ absolute density measurements were carried out using planar LIF.

### B. Rayleigh calibration

The planar LIF diagnostic was calibrated by Rayleigh scattering (RS) in air. The main objectives of this calibration were:

a) To obtain the energy distribution profile of the sheet laser beam; and
b) To obtain all the terms depending on the probe volume, parameters of optical elements used in signal acquisition and the detector efficiency, *etc.*, required in Eq. 4 to obtain quantitative planar LIF measurements.

In the RS calibration of the planar LIF system, the alignment of the beam, laser energy per pulse, voltage applied to the photo-multiplier tube, and iCCD camera settings were kept identical to those used for the LIF measurements in the arc. Since the RS process is less efficient than resonance photon absorption, the RS signals were small and a number of RS images were accumulated to achieve the desired signal-to-noise ratio (SNR). The background was then subtracted to obtain a 2D distribution of the RS signal as schematically shown in Figure 6. The bright spots seen in Figure 6(a) and (c) and originated from impurities in the air were removed in post-processing.

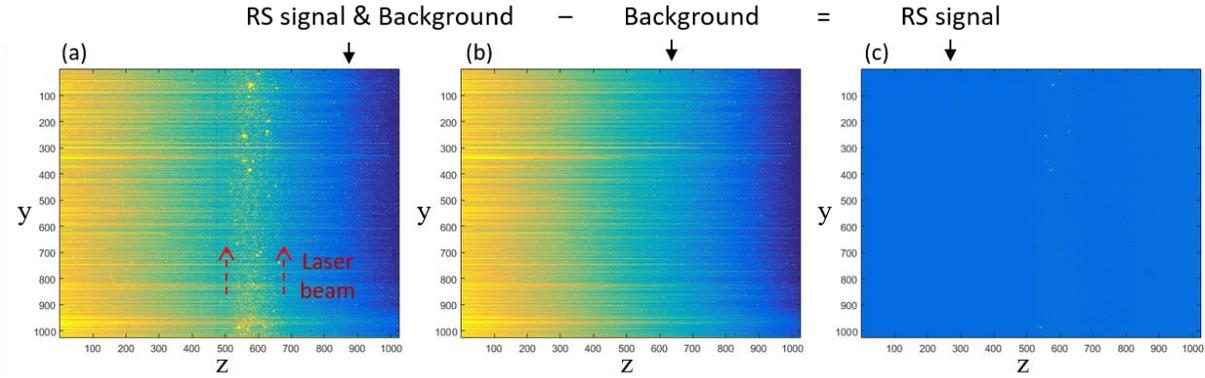

Figure 6. Example of recorded raw image of the RS signal in air with background (a), recorded background (b) and the RS signal (c) resulting from subtraction of (b) from (a).

The RS signal distribution in the transverse direction of the sheet laser beam was obtained by summing the camera pixel intensities along the beam propagation direction (vertical direction in Figure 6) as shown in Figure 7. Since the RS signal is proportional to laser energy, $S_{RS} \propto E_L$, the obtained distribution was used to calculate the normalized laser beam energy distribution profile in the transverse direction (z) as

$$\rho_L(z) = \frac{\sum_y S_{RS}(y,z)}{\sum_{y,z} S_{RS}(y,z)} \qquad (12)$$

where $y$ and $z$ are the coordinates of the individual pixels. Thus, the actual laser energy distribution in the RS measurements can be calculated as $E_L^{RS}(z) = \rho_L(z)\bar{E}_L^{RS}$ with $\bar{E}_L^{RS}$ being the total laser energy per pulse measured by a pyroelectric detector.

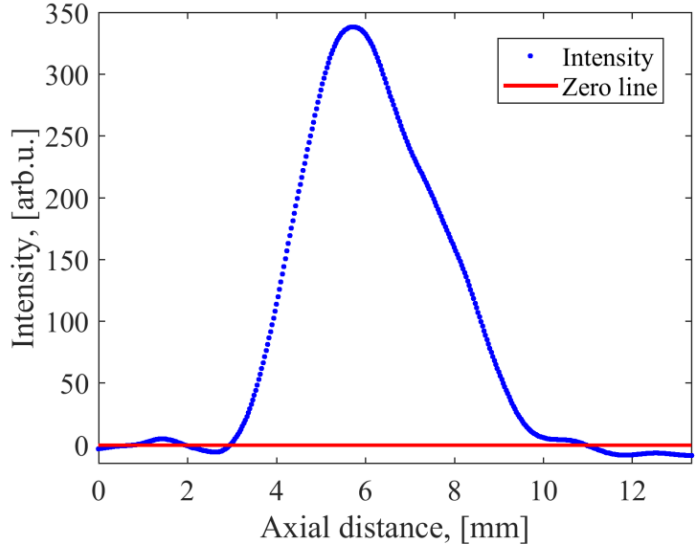

Figure 7. Transverse energy distribution within the planar LIF laser beam.

The close proximity of the LIF excitation and fluorescence wavelengths allows the same profile to be used for LIF signal measurements:

$$E_L^{LIF}(z) = \rho_L(z)\bar{E}_L^{LIF} \tag{13}$$

The RS signal is also proportional to the laser fluence, $I_L = \frac{\bar{E}_L^{RS}}{A_L}$, where $\bar{E}_L^{RS}$ is the laser energy per pulse and $A_L$ is the laser beam cross-sectional area. It can be expressed similarly to Eq. 3:

$$S_{RS} = n_{gas} V \frac{\lambda_L}{hc} \frac{d\sigma_{RS}}{d\Omega} \Omega \, K \, I_L \tag{14}$$

Here, V is the interaction volume, $n_{gas}$ is the particle density of the scatterers, $\Omega$ is the light collection solid angle, K is the cumulative constant including response function and efficiency of the acquisition system. The RS cross-section, $4\pi \frac{d\sigma_{RS}}{d\Omega} = 1.126 \cdot 10^{-26}$ cm², was obtained from [31]. The density of scatterers was calculated from the ideal gas law, $n_{gas} = \frac{p_{gas}}{k_B T_{gas}}$, where $k_B$ is the Boltzmann constant, $p_{gas}$ is the gas pressure, and $T_{gas} = 298\,K$ is the gas temperature. Thus, the RS signal can be written as

$$S_{RS} = \alpha \, \bar{E}_L^{RS} \, p_{gas}, \alpha = \frac{\lambda_L}{hc} \frac{1}{A_L} \frac{d\sigma_{RS}}{d\Omega} \frac{1}{k_B T_{gas}} V \, \Omega \, K \tag{15}$$

where α can be obtained from a linear fit to the RS signal as a function of gas pressure and laser energy, as shown in Figure 8. It includes most of unknown terms in Eq. 4 except for the effective lifetime of the excited state, which is discussed in the following section.

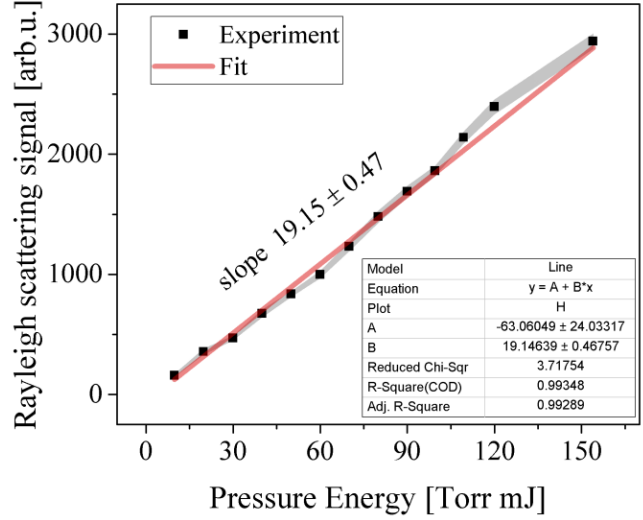

Figure 8. Dependence of the RS signal on product of the gas pressure and laser energy for planar LIF geometry. The gray band represents the measurement errors. The slope α defines the unknown terms in Eq. 4 used for quantitative planar LIF measurements.

### C. Lifetime of the $C_2$ excited state

The apparent lifetime of the $C_2$ excited state in an atmospheric pressure plasma strongly depends on the radiation lifetime and quenching rates. Additional terms related to dissociation and coagulation processes leading to the loss of carbon dimers can also shorten the excited state lifetime. The effective lifetime can be written in general form as

$$\tau_{eff} = [A + \sum_i n_i\, q_i + O]^{-1} \qquad (16)$$

where each term is temperature dependent and some terms may depend on the local density of quenchers assuming constant background gas pressure. Here, A is the spontaneous emission coefficient, (corresponding lifetime of 120 ns [32]) and $n_i$ is the quencher number density for quenchers of type i with quenching coefficient $q_i$. The last term, O, represents all other processes leading to reduction of the lifetime of the state. When quenching rates are known for certain quenchers in gas mixtures, their dependence on temperature and number density complicates the use of Eq. 16 in most cases when non-stationary atmospheric plasma is involved as is the case for the carbon arc [26,27].

The first measurements of the LIF signal decay revealed that the effective lifetime of the excited state is comparable to the laser pulse duration, 8 ns, as shown in Figure 9. However, after the extinction of the laser pulse, the rate equation for the excited state is reduced to a decay problem with a single-exponential solution characterized by the effective life time $\tau_{eff}$:

$$\frac{dn_{exc}}{dt} = -\frac{n_{exc}}{\tau_{eff}} \qquad (17)$$

where $n_{exc}$ is the population of the excited state.

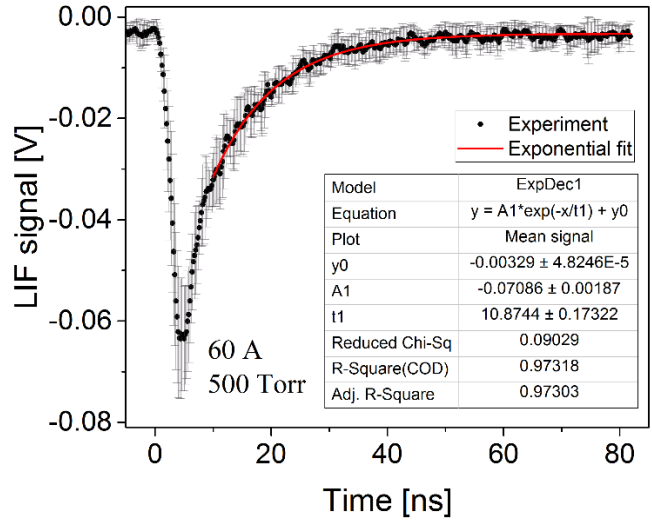

Figure 9. The measured LIF signal decay.

Therefore, the decay time of the LIF signal was measured starting 10 ns after the extinction of the laser pulse beginning with the lower background gas pressure for which the decay time is longer. The pressure was gradually increased to 500 Torr keeping all other parameters identical, e.g., discharge current and voltage, PMT bias voltage, laser beam energy, *etc.* Since the helium is a viscous medium preventing expansion of the arc plasma, the pressure decrease led to expansion of the $C_2$ bubble. Thus, the detection window was scanned to find the maximum $C_2$ density at each pressure value and the corresponding decay time was measured. The obtained pressure dependence of the decay time is shown in Figure 10.

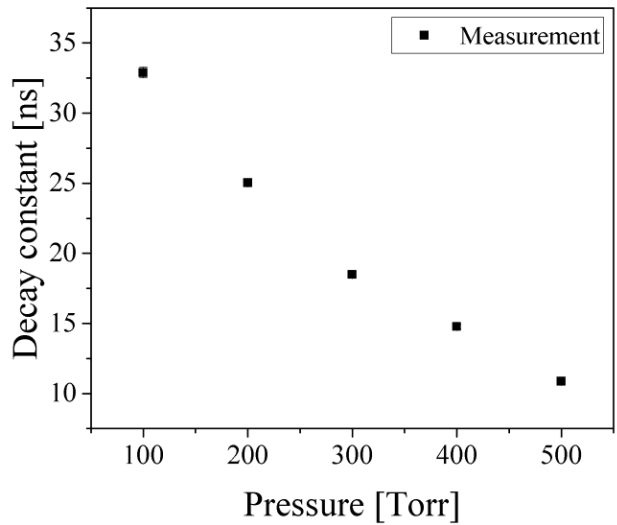

Figure 10. Effective lifetime of the excited state of $C_2$ as a function of background gas pressure.

### D. Saturation effect

Compared with conventional LIF, the planar LIF geometry makes possible instantaneous measurements of fluorescence over a large area with high spatial and temporal resolution. This is valuable for measurements in the arc discharge for nanomaterial synthesis, which is characterized by fast shape variation and unpredictable motion [26,27]. However, this implies a strong requirement for single shot measurements by planar LIF. Since spontaneous emission is the main contributor to the background level in the planar LIF measurements, the choice of $C_2$ excitation scheme (3,1) was the one with the largest signal-to-noise (SNR) ratio. Transitions with low vibrational quantum numbers such as, (0,0), (0,1), etc., are more intense, meaning that their upper states are well populated and thus effect of the population increase due to stimulated absorption of laser photons may not be large.

Another option is to increase the laser energy since the LIF signal is proportional to the laser energy. This applies until the contribution of laser-stimulated emission is small and does not affect the population of the excited state. In general, this effect of transition saturation can be accounted for by the saturation parameter s, which affects the LIF signal according to:

$$S_{LIF} = Y_{LIF} \frac{E_L}{1+sE_L} \tag{18}$$

where $E_L$ is the laser energy, $S_{LIF}$ is the measured LIF signal and $Y_{LIF}$ is the LIF signal per unit laser energy without the saturation effect. For the planar LIF geometry the saturation parameter becomes a function of transversal beam coordinate following the spatial profile of the laser intensity shown in Figure 7. However, it was impossible to derive such a dependence due to large error bars. Therefore, the saturation parameter was calculated from the fit of the calibration curve shown in Figure 11 with integrated LIF signal. The measured LIF signal was then corrected using the obtained parameter $s = 0.036 \pm 0.01$ (1/µJ) and $Y_{LIF} = 3.00 \pm 1.15$ (1/µJ).

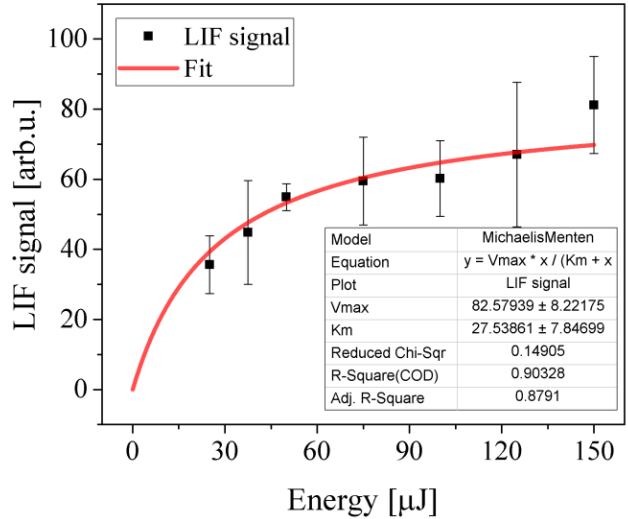

Figure 11. Saturation curve for the planar LIF diagnostic. The fit provides the saturation parameter s. The last point of the experimental curve was excluded from the fit.